\def\year{2018}\relax
\documentclass[letterpaper]{article} %DO NOT CHANGE THIS
\usepackage{aaai18}  %Required
\usepackage{times}  %Required
\usepackage{helvet}  %Required
\usepackage{courier}  %Required
\usepackage{url}  %Required
\usepackage{graphicx}  %Required
\usepackage{amssymb}
\usepackage{amsfonts}
\usepackage{amsmath}
\usepackage{amsthm}
\usepackage{subfig}
\usepackage{comment}
\usepackage{paralist}
\usepackage{multirow}

\newcommand{\figref}[1]{Fig.~\ref{fig:#1}}

\theoremstyle{definition}

\newtheorem{definition}{Definition}

\DeclareMathOperator*{\argmin}{argmin}

\title{Human-in-the-Loop SLAM}
\author{Samer B. Nashed \and Joydeep Biswas \\
  College of Information and Computer Sciences\\
  140 Governors Drive\\
  Amherst MA, USA 01003
}

\begin{document}

\maketitle

\begin{comment}
TODO:

double check style

minimal .tex + .bib + .bbl + images bundle

\end{comment}

\begin{abstract}
Building large-scale, globally consistent maps is a challenging problem, made more
 difficult in environments with limited access, sparse
features, or when using data collected by novice users. For such scenarios,
where state-of-the-art mapping algorithms produce globally inconsistent maps, we
introduce a systematic approach to
incorporating sparse human corrections, which we term Human-in-the-Loop
Simultaneous Localization and Mapping (HitL-SLAM). Given an initial
factor graph for pose graph SLAM, HitL-SLAM accepts approximate, potentially
erroneous, and rank-deficient human input, infers the intended correction via
expectation maximization (EM), back-propagates the extracted corrections over
the pose graph, and finally jointly optimizes the factor graph including
the human inputs as human correction factor terms, to yield globally consistent large-scale maps.
We thus contribute an EM formulation for inferring potentially
rank-deficient human corrections to mapping, and human correction factor extensions to the
factor graphs for pose graph SLAM that result in a principled approach to joint
optimization of the pose graph while simultaneously accounting for multiple
forms of human correction. We present empirical results showing the
effectiveness of HitL-SLAM at generating globally accurate and consistent maps
even when given poor initial estimates of the map.
\end{abstract}

\section{Introduction}

Building large-scale globally consistent metric maps requires accurate relative
location information between poses with large spatial separation. However, due
to sensor noise and range limitations, such correlations across distant poses
are difficult to extract from real robot sensors. Even when such observations
are made, extracting such correlations autonomously is a computationally
intensive problem. Furthermore, the order of exploration, and the speed of the
robot during the exploration affect the numerical stability, and consequently
the global consistency of large-scale maps. Due to these factors, even state of
the art mapping algorithms often yield inaccurate or inconsistent large-scale
maps, especially when processing data collected by novice users in challenging
environments.

\begin{figure}[tb]
  \centering
  \includegraphics[scale=0.103]{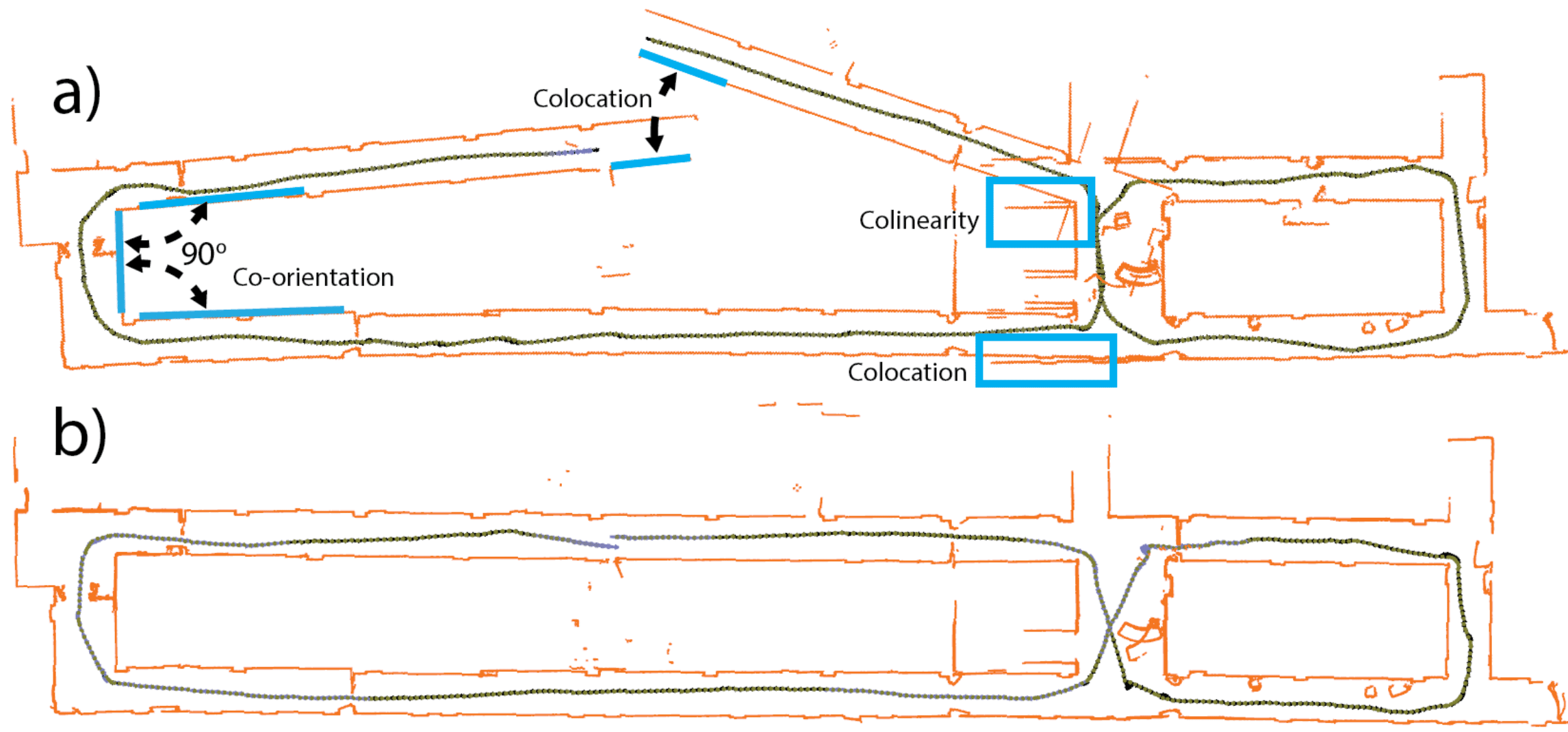}
  \caption{HitL SLAM example, showing a) the input initial map with global consistency
  errors, and b) the resulting final map produced by HitL-SLAM by incorporating human
  corrections (blue lines) along with the input.}
\vspace{-1.5em}
  \label{fig:teaser}
\end{figure}

To address these challenges and limitations of large-scale mapping, we propose
Human-in-the-Loop SLAM (HitL-SLAM), a principled approach to incorporate
approximate human corrections in the process of solving for metric
maps\footnote{Code and sample data is available at:\\ \url{https://github.com/umass-amrl/hitl-slam}}.
\figref{teaser} presents an example of HitL-SLAM in practice. HitL-SLAM operates
on a pose graph estimate of a map along with the corresponding observations from each
pose, either from an existing state-of-the-art SLAM solver, or aligned only by
odometry. In an interactive, iterative process, HitL-SLAM accepts human
corrections, re-solves the pose graph problem, and presents the updated map
estimate.  This iterative procedure is repeated until the user is satisfied by
the mapping result, and provides no further corrections.

We thus present three primary contributions:
\begin{inparaenum}[1)]
  \item an EM-based algorithm~\cite{EM} to interpret several types of approximate human
  correction for mapping (\S 4),
  \item a human factor formulation to incorporate a variety of types of human corrections
  in a factor graph for SLAM (\S 5), and
  \item a two-stage solver for the resultant, hybrid factor graph composed of human and robot
   factors, which minimally distorts trajectories in the presence of rank deficient human corrections (\S 5).
\end{inparaenum}
We show how HitL-SLAM introduces numerical stability in the
mapping problem by using human corrections to introduce off-diagonal blocks in the information matrix.
Finally, we present several examples of HitL-SLAM operating on maps that
intrinsically included erroneous observations and poor initial map estimates, and producing accurate, globally consistent maps.

\section{Related Work}

Solutions to robotic mapping and SLAM have improved dramatically in recent
years, but state-of-the-art algorithms still fall short at being able to
repeatable and robustly produce globally consistent maps,
particularly when deployed over large areas and by non-expert users. This is in
part due to the difficulty of the data association problem
~\cite{dissanayake2011review,bailey2006simultaneous,aulinas2008slam}. The idea
of humans and robots collaborating in the process of map building to overcome
such limitations is not new, and is known as Human-Augmented Mapping (HAM).

Work within HAM belongs primarily to one of two groups, depending on whether
the human and robot collaborate in-person during data collection (C-HAM),
or whether the human provides input remotely or after the data collection
(R-HAM). Many C-HAM techniques exist to address semantic
~\cite{nieto2010semantic,christensen2010detecting} and topological
~\cite{topp2006topological} mapping. A number of approaches have also been
proposed for integrating semantic and topological information, along with human
trackers ~\cite{milella2007laser}, interaction models ~\cite{topp2006bringing},
and appearance information~\cite{pronobis2012large}, into \emph{conceptual
spatial maps} ~\cite{zender2007integrated}, which are organized in a
hierarchical manner.

There are two limitations in these C-HAM approaches. First, a
human must be present with the robot during data collection. This places
physical constraints on the type of environments which can be mapped, as they
must be accessible and traversable by a human. Second, these methods are
inefficient with respect to the human's attention, since most of the time the
human's presence is not critical to the robot's function, for instance during
navigation between waypoints. These approaches, which focus mostly on semantic
and topological mapping, also typically assume that the robot is able to
construct a nearly perfect metric map entirely autonomously. While this is
reasonable for small environments, globally consistent metric mapping of large,
dynamic spaces is still a hard problem.

In contrast, most of the effort in R-HAM has been concentrated on either
incorporating human input remotely via tele-operation such as in the Urban
Search and Rescue (USAR) problem ~\cite{murphy2004human,nourbakhsh2005human}, or
in high level decision making such as goal assignment or coordination of
multiple agents
~\cite{olson2013exploration,parasuraman2007adaptive,doroodgar2010search}. Some
R-HAM techniques for metric mapping and pose estimation have also been explored,
but these involve either having the robot retrace its steps to fill in parts
missed by the human ~\cite{kim2009human} or by having additional agents and
sensors in the environment ~\cite{kleiner2007mapping}.

A number of other approaches have dealt with interpreting graphical or textual human input within the contexts of localization~\cite{behzadian2015monte,boniardi2016autonomous} and semantic mapping~\cite{hemachandra2014information}. While these approaches solve similar signal interpretation problems, this paper specifically focuses on metric mapping.

Ideally, a robot could explore an area only once with no need for human guidance
or input during deployment, and later with minimal effort, a human could make
any corrections necessary to achieve a near-perfect metric map. This is
precisely what HitL-SLAM does, and additionally HitL-SLAM does not
require in-person interactions between the human and robot during the data
collection.

\section{Human-in-the-Loop SLAM}

HitL-SLAM operates on a factor graph $G = \{X, F \}$, where $X$ is the set of estimated
poses along the robot's trajectory, and $F = \{R, H \}$ is the set of factors which encode
information about both relative pose constraints arising from odometry and observations,
$R$, and constraints supplied by the human, $H$. The initial factor graph $G_0$ may be
provided by any pose graph SLAM algorithm, and HitL-SLAM is capable of handling
constraints in $G_0$ with or without loop closure. In our experiments, we used Episodic
non-Markov Localization (EnML)~\cite{biswas2017episodic}  without any explicit loop
closures beyond the length of each episode.

HitL-SLAM runs iteratively, with the human specifying constraints on observations in the
map, and the robot then enforcing those constraints along with all previous constraints to
produce a revised estimate of the map. To account for inaccuracies in human-provided
corrections, interpretation of the such input is necessary before human correction factors
can be computed and added to the factor graph. Each iteration, the robot first proposes an
initial graph $G_i = \{ X_i, F_i \}$, then the human supplies a set of correction factors
$H_i$, and finally the robot re-optimizes the poses in the factor graph, producing $G_i' =
\{ X_i', F_i \cup H_i \}$.

\begin{definition}
A {\bf human correction factor}, \emph{h}, is defined by the tuple $h=\langle
P_a, P_b, S_a, S_b, X_a, X_b, m \rangle$ with:

\enumerate

\item[$\bullet$ $P_a, P_b \subset \mathbb{R}^2$] : Sets of end-points of the
two line segments $a,b$ drawn by the human,

\item[$\bullet$ $S_a, S_b \subset S$] : Sets of observations selected by the
two line segments $a, b$ respectively,

\item[$\bullet$ $X_a, X_b \subset X_{1:t}$] : Sets of poses from which the
observations $S_a,S_b$ were made,

\item[$\bullet$ $m \in M$] : The mode of correction.

\end{definition}

\begin{figure}[htb]
  \centering
  \includegraphics[scale=0.075]{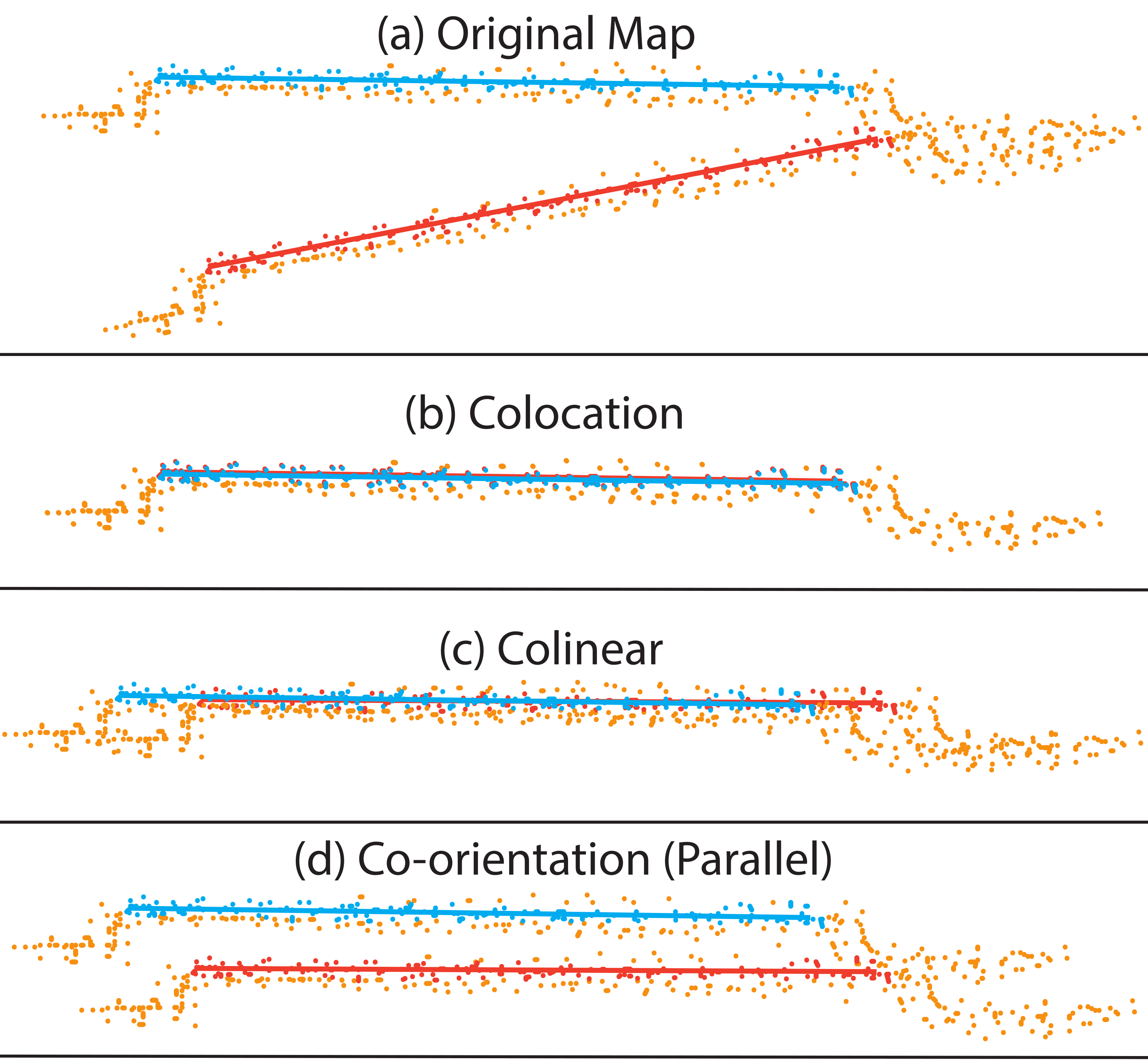}
  \caption{Result of transforming observation point clouds based on different
human constraints, showing (a) Original map, (b) Colocation constraint, (c)
Collinear constraint, (d) Co-orientation constraint. In all sub-figures the red
and blue lines denote $P_a$ and $P_b$, respectively, and red and blue points
denote $S_a$ and $S_b$. $S \setminus (S_a \cup S_b)$ appear in orange.}
\label{fig:corr}
\end{figure}

$S_a, S_b$ are subsets of all observations $S$, and poses $x_i$ are added to
the sets $X_a, X_b$ if there are observations in $S_a, S_b$ that arising from
pose $x_i$. $M$ is an enumeration of the modes of human correction, a subset of which are shown in \figref{corr}. The modes $M$ of correction are defined as follows:

\begin{enumerate}
\item \emph{Colocation:} A full rank constraint specifying that two sets of
observations are at the same location, and with the same orientation.

\item \emph{Collinearity:} A rank deficient constraint specifying that two sets of
observations are on the same line, with an unspecified translation along the
line.

\item \emph{Perpendicularity:} A rank deficient constraint specifying that the
two sets of observations are perpendicular, with an unspecified translation
along either of their lines.

\item \emph{Parallelism:}  A rank deficient constraint specifying that the
two sets of observations are parallel, with an unspecified translation
along the parallel lines.

\end{enumerate}

\begin{figure}[htb]
  \centering
  \includegraphics[scale=0.09]{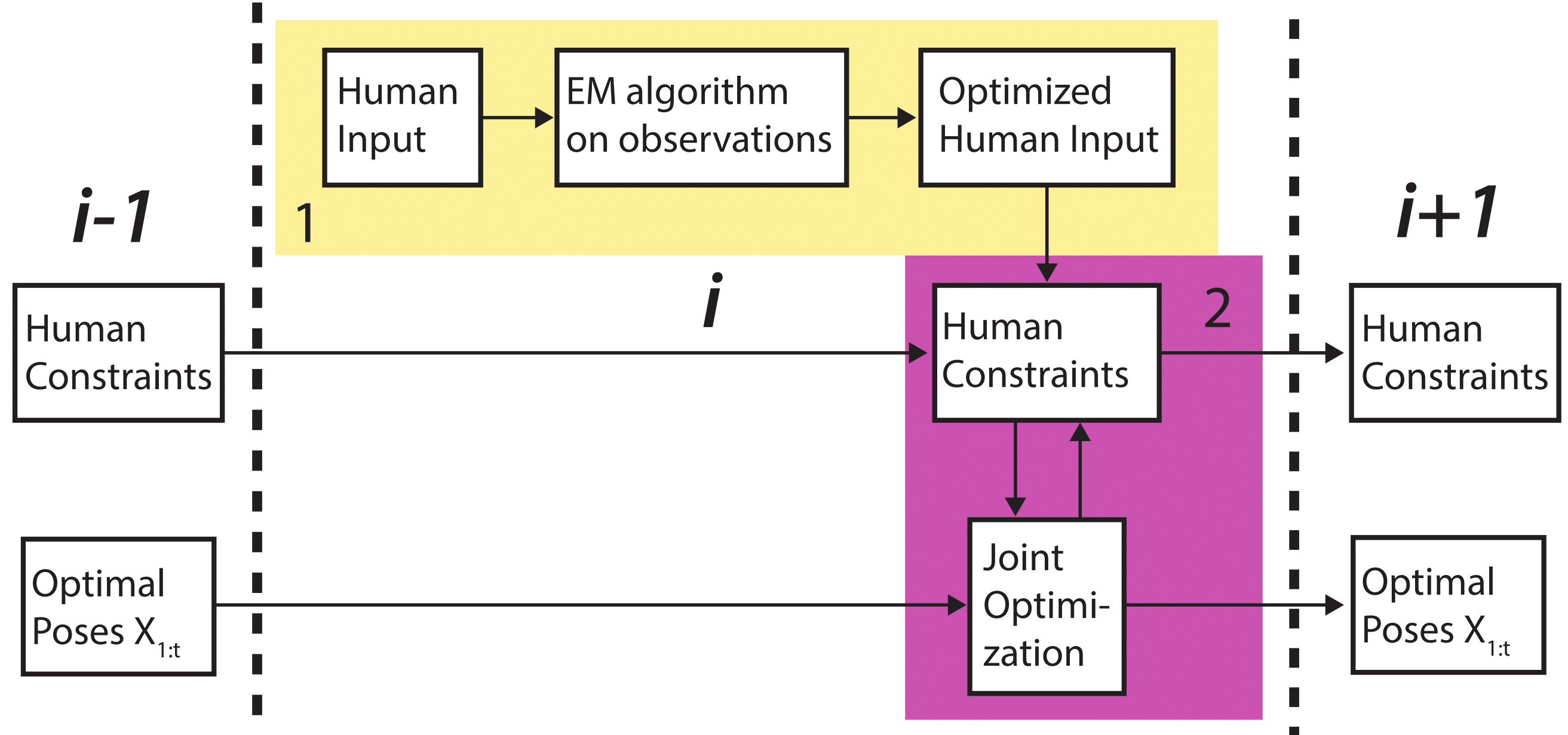}
  \caption{Flow of information during processing of the $i^{\text{th}}$ human
input. Block 1 (yellow) outlines the evaluation of human input, and block 2
(purple) outlines the factor graph construction and optimization processes.
Note that the joint optimization process optimizes $\textit{both}$ pose
parameters $\textit{and}$ human constraint parameters.}
\label{fig:BlockDiag}
\end{figure}

Each iteration of HitL-SLAM proceeds in two steps, shown in \figref{BlockDiag}.
First, the human input is gathered, interpreted, and a set of human correction
factors are instantiated (Block 1). Second, a combination of analytical and
numerical techniques is used to jointly optimize the factor graph using both
the human correction factors and the relative pose factors (Block 2). The
resulting final map may be further revised and compressed by algorithms such as
Long-Term Vector Mapping~\cite{nashed2016curating}.

We model the problem of interpreting human input as finding the observation
subsets $S_a, S_b$ and human input sets $P_a, P_b$ which maximize the joint
correction input likelihood, $p(S_a, S_b, P_a, P_b | P_a ^0, P_b^0, m)$, which
is the likelihood of selecting observation sets $S_a, S_b$ and point sets $P_a,
P_b$, given initial human input  $P_a ^0, P_b^0$ and correction mode $m$. To
find $S_a, S_b$ and $P_a, P_b$ we use the sets $P_a ^0, P_b ^0$ and observations
in a neighborhood around $P_a ^0, P_b ^0$ as initial estimates in an Expectation
Maximization approach. As the pose parameters are adjusted during optimization
in later iterations of HitL-SLAM, the locations of points in $P_a, P_b$ may
change, but once an observation is established as a member of $S_a$ or $S_b$ its
status is not changed.

Once $P_a, P_b$ and $S_a, S_b$ are determined for a new constraint, then given $m$ we can find the set of poses $X^* _{1:t}$ which best satisfy all given constraints. We first compute an initial estimate $X^0 _{1:t}$ by analytic back-propagation of the most recent human correction factor, considering sequential constraints in the pose-graph. Next, we construct and solve a joint optimization problem over the relative pose factors $r$ and the human correction factors $h$. This amounts to finding the set of poses $X^*_{1:t}$ which minimize the sum of the cost of all factors,

\begin{equation*}
X^*_{1:t} = \argmin_{X_{1:t}}\left[\sum_{i=1}^{|R|} c_r (r_i)+\sum_{j=1}^{|H|} c_m (h_j)\right],
\end{equation*}

\noindent where $c_r: R \rightarrow \mathbb{R}$ computes the cost from relative pose-graph factor $r_i$, and $c_m: H \rightarrow \mathbb{R}$ computes the cost from human correction factor $h_j$ with correction mode $m$. Later sections cover the construction of the human correction factors and the formulation of the optimization problem.

\section{Interpreting Human Input}

\subsection{Human Input Interface}

HitL-SLAM users provide input by first entering the `provide correction' state by pressing the `p' key. Once in the `provide correction' state, they enter points for $P_a$, $P_b$ by clicking and dragging along the feature (line segment) they wish to specify. The mode $m$ is determined by which key is held down during the click and drag. For instance, CTRL signifies a colocation constraint, while SHIFT signifies a collinear constraint. To finalize their entry, the user exits the `provide correction' state by again pressing the `p' key. Exit from the `provide correction' state triggers the algorithm in full, and the user may specify additional corrections once a revised version of the map is presented.

\subsection{Human Input Interpretation}

Due to a number of factors including imprecise input devices, screen
resolution, and human error, what the human actually enters and what they
intend to enter may differ slightly. Given the raw human input line segment
end-points $P_a ^0,P_b ^0$ and the mode of correction $m$, we frame the
interpretation of human input as the problem of identifying the observation
sets $S_a, S_b$ and the effective line segment end-points $P_a, P_b$ most
likely to be captured by the potentially noisy points $P_a ^0,P_b ^0$.
 To do this we use the EM algorithm, which maximizes the
log-likelihood $\ell$,

\begin{equation*}
\ell(\theta) = \sum_i \sum_{z_i} p(z_i | s_i , \theta^{\text{old}}) \log(p(z_i , s_i | \theta )),
\end{equation*}

\noindent where the parameters $\theta = \{ P_a, P_b \}$ are the
interpreted human input (initially assumed to be  $P_a ^0,P_b ^0$), the $s_i
\in S$ are the observations, and the latent variables $z_i$ are indicator
variables denoting the inclusion or exclusion of $s_i$ from $S_a$ or $S_b$. The
expressions for $p(z_i | s_i , \theta^{\text{old}})$ and $p(z_i , s_i | \theta
)$ come from a generative model of human error based on the normal
distribution, $\mathcal{N}(\mu(\theta), \sigma^2)$. Here, $\sigma$ is the
standard deviation of the human's accuracy when manually specifying points, and
is determined empirically; $\mu(\theta)$ is the center or surface of the
feature.

Let $\delta(s_i, \theta)$ be the squared Euclidean distance between a given
observation $s_i$ and the feature (in this case a line segment) parameterized by
$\theta$. Note that $p(z_i | s_i, \theta)$ is convex due to our Gaussian model
of human error. Thus, the EM formulation reduces to iterative least-squares over
changing subsets of $S$ within the neighborhoods of $P_a, P_b$. The raw human
inputs $P_a ^0,P_b ^0$ are taken as the initial guess to the solution $\theta$,
and are successively refined of iterations of the EM algorithm to compute the
final interpreted human input  $P_a,P_b$.

Once $P_a, P_b$ have been determined, along with observations $S_a, S_b$, we can
find the poses responsible for those observations $X_a, X_b$, thus fully
defining the human correction factor $h$. To make this process more robust to
human error when providing corrections, a given pose is only allowed in $X_a$ or
$X_b$ if there exist a minimum of $T_p$ elements in $S_a$ or $S_b$ corresponding
to that pose. The threshold $T_p$ is used for outlier rejection of provided
human corrections. It is empirically determined by evaluating a human's ability
to accurately select points corresponding to map features, and is the minimum
number of points a feature must have for it to be capable of being repeatedly
and accurately selected by a human.

\section{Solving HitL-SLAM}

After interpreting human input, new pose estimates are computed in three steps.
First, all explicit corrections indicated by the human are made by applying the
appropriate transformation to $X_b$ and subsequent poses. Next, any resultant
discontinuities are addressed using Closed-Form Online Pose-Chain SLAM
(COP-SLAM)~\cite{copslam}. And last, final pose parameters are calculated via
non-linear least-squares optimization of a factor graph. The three-step approach
is necessary in order to avoid local minima.

\subsection{Applying Explicit Human Corrections}

Although the user may select sets of observations in any order, we define all poses $x_i \in X_a$ to occur before all poses $x_j \in X_b$. That is, $P_a$ is the input which selects observations $S_a$ arising from poses $X_a$ such that $\forall x_i \in X_a$ and $x_j \in X_b$, $i < j$, where $X_b$ is defined analogously by observations $S_b$ specified by input $P_b$.

Given $P_a$ and $P_b$, we find the affine correction transformation $A$ which
transforms the set of points defined by $P_b$ to the correct location relative
to the set of points defined by $P_a$, as specified by mode $m$. If the
correction mode is rank deficient, we force the motion of the observations as a
whole to be zero along the null space dimensions. For co-orientation, this means
that the translation correction components of $A$ are zero, and for
collinearity the translation along the axis of collinearity is zero.
\figref{corr} shows the effect of applying different types of constraints to a
set of point clouds.

After finding $A$ we then consider the poses in $X_b$ to constitute points on a rigid body, and transform that body by $A$. The poses $x_k$ such that $\forall x_j \in X_b$, $k > j$, are treated similarly, such that the relative transformations between all poses occurring during or after $X_b$ remain unchanged.

\subsection{Error Backpropagation}

If $X_a \cup X_b$ does not form a contiguous sequence of poses, then this explicit change creates at least one discontinuity between the earliest pose in $X_b$, $x^b _0$ and its predecessor, $x_c$. We define affine transformation $C$ such that $x^b _0 = A_{cb} C x_c$, where $A_{cb}$ was the original relative transformation between $x_c$ and $x_0 ^b$. Given $C$, and the pose and covariance estimates for poses between $X_a$ and $X_b$, we use COP-SLAM over these intermediate poses to transform $x_c$ without inducing further discontinuities.

The idea behind COP-SLAM is a covariance-aware distribution of translation and
rotation across many poses, such that the final pose in the pose-chain ends up
at the correct location and orientation. The goal is to find a set of updates
$U$ to the relative transformations between poses in the pose-chain such that $C
= \prod_{i=1} ^n U_i$.

COP-SLAM has two primary weaknesses as a solution to applying human
corrections in HitL-SLAM. First, it requires translation uncertainty estimates
to be isotropic, which is not true in general. Second, COP-SLAM deals poorly
with nested loops, where it initially produces good pose estimates but during
later adjustments may produce inconsistencies between observations. This is
because COP-SLAM is not able to simultaneously satisfy both current and
previous constraints. Due to these issues, we use COP-SLAM as an initial
estimate to a non-linear least-squares optimization problem, which produces a
more robust, globally consistent map.

\subsection{HitL-SLAM Optimization}
\label{sec:hitl-optimization}
Without loop closure, a pose-chain of $N$ poses has $\mathcal{O}(N)$ factors.
With most loop closure schemes, each loop can be closed by adding one
additional factor per loop. In HitL-SLAM, the data provided by the human is
richer than most front-end systems, and reflecting this in the factor graph
could potentially lead to a prohibitively large number of factors. If $|X_a| =
n$ and $|X_b| = m$, then a na\"ive algorithm that adds a factor between all
pairs $(x^a _i, x^a_j)$, $(x^a_i, x^b _j)$, and $(x^b_i, x^b_j)$, where $x^a \in
X_a$ and $x^b \in X_b$, would add $(m+n)^2$ factors for \emph{every loop}. This
is a poor approach for two reasons. One, the large number of factors can slow
down the optimizer and potentially prevent it from reaching the global optimum.
And two, this formulation implies that every factor is independent of every
other factor, which is incorrect.

Thus, we propose a method for reasoning about human correction factors jointly,
in a manner which creates a constant number of factors per loop while also
preserving the structure and information of the input. Given a human correction
factor $h = \langle P_a, P_b, S_a, S_b, X_a, X_b, m \rangle$, we define $c_m$ as
the sum of three residuals, $R_a$, $R_b$, and $R_p$. The definitions of $R_a$
and $R_b$ are the same regardless of the correction mode $m$:
\begin{equation*}
R_a = \left ( \frac{ \sum_{i=1} ^{|S_a|} \delta (s_i ^a , P_a)}{|S_a|} \right ) ^{\frac{1}{2}}, \hspace{0.5mm} R_b = \left ( \frac{ \sum_{i=1} ^{|S_b|} \delta (s_i ^b , P_b)}{|S_b|} \right ) ^{\frac{1}{2}}.
\end{equation*}

\noindent As before, $\delta(s, P)$ denotes the squared Euclidean distance from observation $s$ to the closest point on the feature defined by the set of points $P$. All features used in this study are line segments, but depending on $m$, more complicated features with different definitions for $\delta(s, P)$ may be used. $R_a$ implicitly enforces the interdependence of different $x_a \in X_a$, since moving a pose away from its desired relative location to other poses in $X_a$ will incur cost due to misaligned observations. The effect on $X_b$ by $R_b$ is analogous.

\begin{figure}[!h]
  \centering
  \includegraphics[scale=0.06]{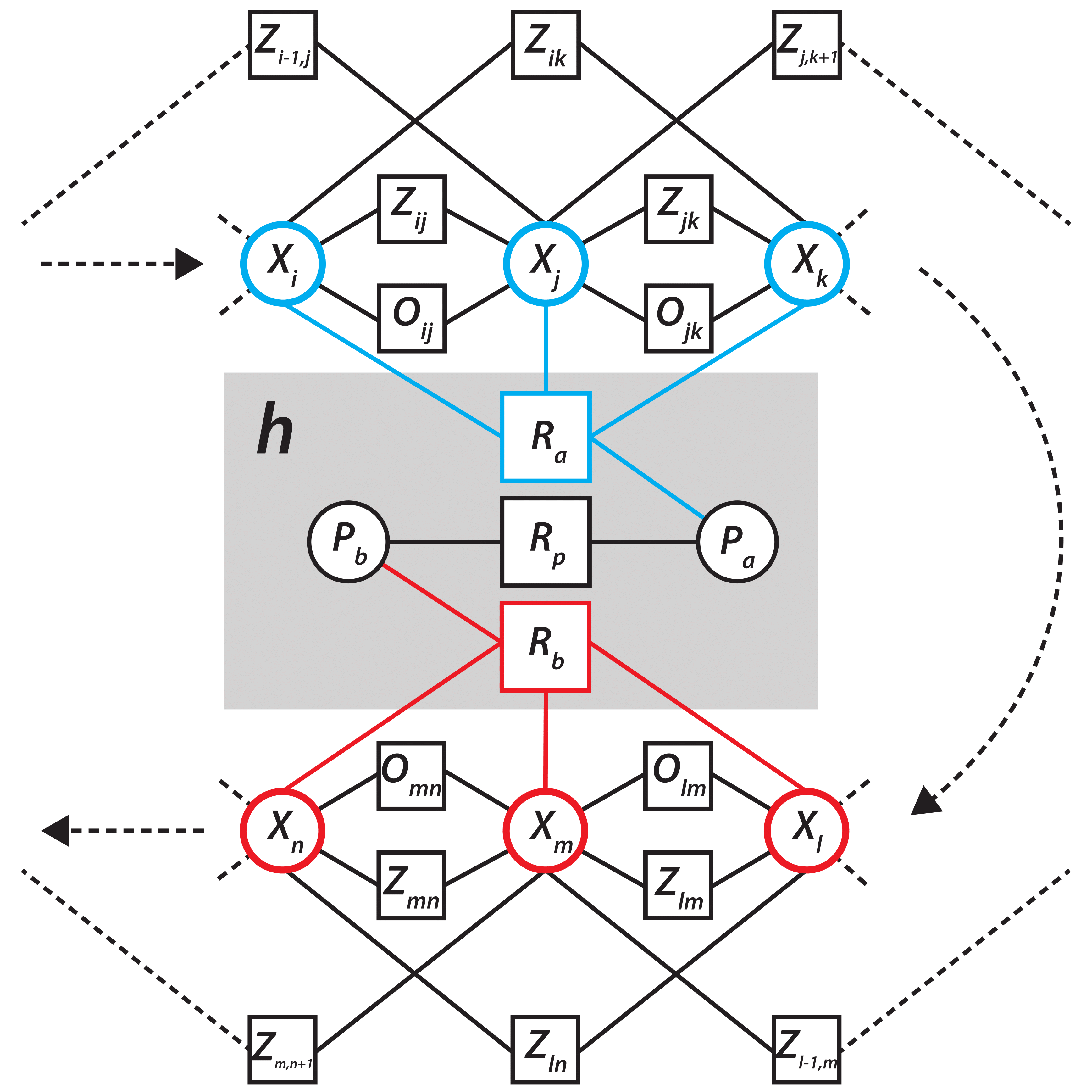}
  \caption{Subset of a factor graph containing a human factor $h$. Factors $R_a$  and $R_b$ drive observations in $S_a$  and $S_b$ toward features $P_a$ and $P_b$, respectively. Factor $R_p$ enforces the geometric relationship between $P_a$ and $P_b$. Note that parameters in $X_a$ (blue poses) and $X_b$ (red poses) as well as $P_a$ and $P_b$ are jointly optimized.}
\label{fig:FG}
\end{figure}

The relative constraints between poses in $X_a$ and poses in $X_b$ are enforced indirectly by the third residual, $R_p$. Depending on the mode, colocation ($+$), collinearity ($-$), co-orientation parallel ($\parallel$), co-orientation perpendicular ($\perp$), the definition changes:

\begin{equation*}
\begin{aligned}
R_p ^+ & = K_1||cm_b - cm_a|| + K_2(1 - (\hat{n}_a \cdot \hat{n}_b)), \\
R_p ^-  & = K_1||(cm_b - cm_a) \cdot \hat{n}_a|| + K_2(1 - (\hat{n}_a \cdot \hat{n}_b)), \\
R_p ^{\parallel} \hspace{0.65mm} & = K_2(1 - (\hat{n}_a \cdot \hat{n}_b)), \\
R_p ^{\perp} & = K_2(\hat{n}_a \cdot \hat{n}_b). \\
\end{aligned}
\end{equation*}

\noindent Here, $cm_a$ and $cm_b$ are the centers of mass of $P_a$ and $P_b$,
respectively, and $\hat{n}_a$ and $\hat{n}_b$ are the unit normal vectors for
the feature (line) defined by $P_a$ and $P_b$, respectively. $K_1$ and $K_2$ are
constants that determine the relative costs of translational error ($K_1$) and
rotational error ($K_2$). The various forms of $R_p$ all drive the points in
$P_b$ to the correct location and orientation relative to $P_a$. During
optimization the solver is allowed to vary pose locations and orientations, and
by doing so the associated observation locations, as well as points in $P_a$ and
$P_b$. \figref{FG} illustrates the topology of the human correction factors in
our factor graph.

\begin{figure}[htb]
  \centering
  \includegraphics[scale=0.106]{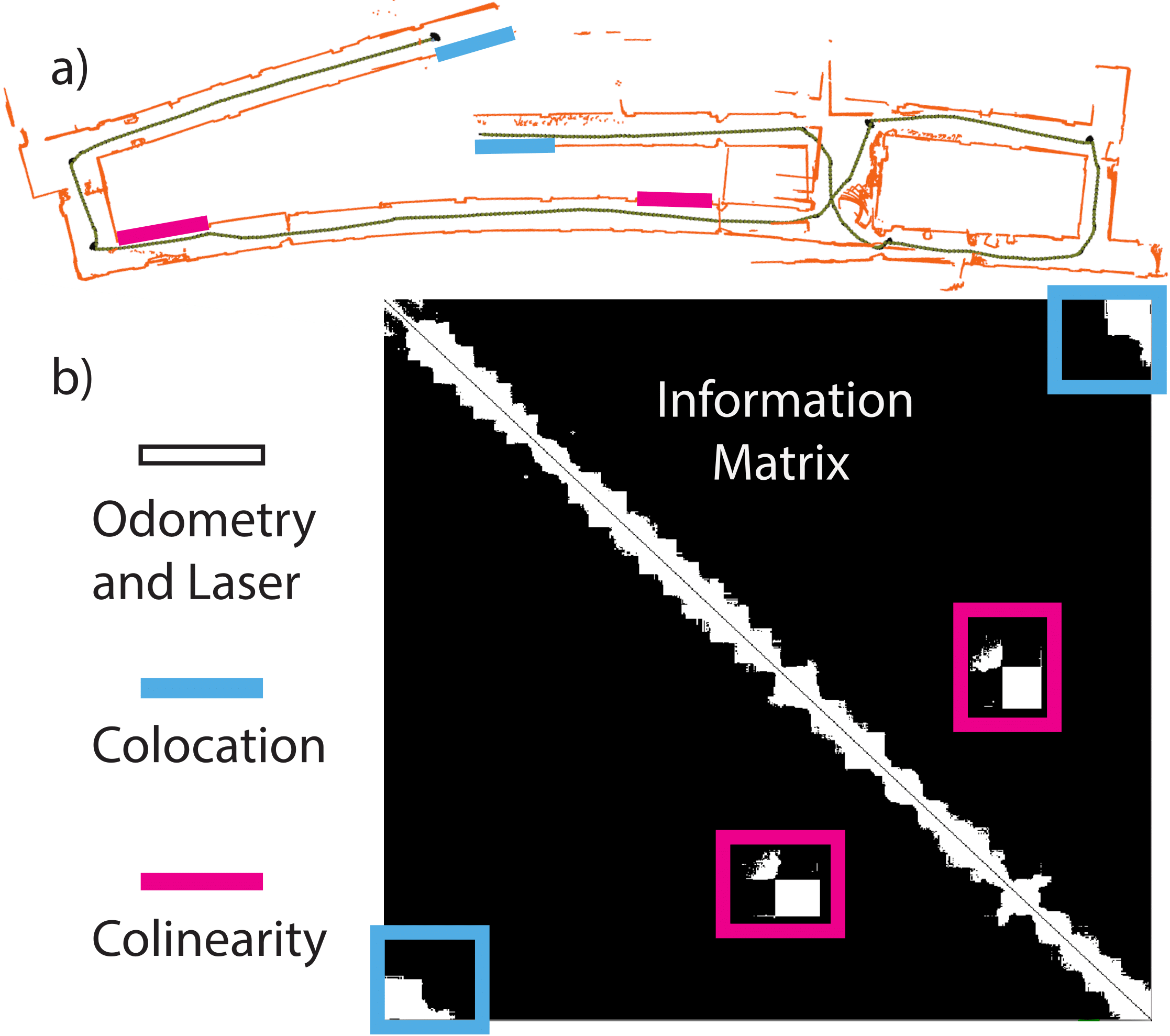}
  \caption{Example map (a) with corrections and resulting information matrix
(b). The white band diagonal represents the correlations from the initial
factor graph $G_0$. The colored lines on the map show the human
correction input: colocation (blue) and collinear (magenta).. The constraints
correspond to the blue and magenta off-diagonal entries in the information
matrix.}
\label{fig:InfoMat}
\end{figure}

Note that HitL-SLAM allows human correction factors to be added to the
factor graph in a larger set of situations compared to autonomous loop closure.
HitL-SLAM introduces `information' loop closure by adding correlations
between distant poses without the poses being at the same location as in
conventional loop closure. The off-diagonal elements in the information
matrix thus introduced by HitL-SLAM assist in enforcing global consistency
just as the off-diagonal elements introduced by loop closure. \figref{InfoMat}
further illustrates this point -- note that the information matrix is still
symmetric and sparse, but with the addition of off-diagonal elements from the
human corrections.

\begin{figure*}[t]
	\centering
	\includegraphics[scale = 0.0385]{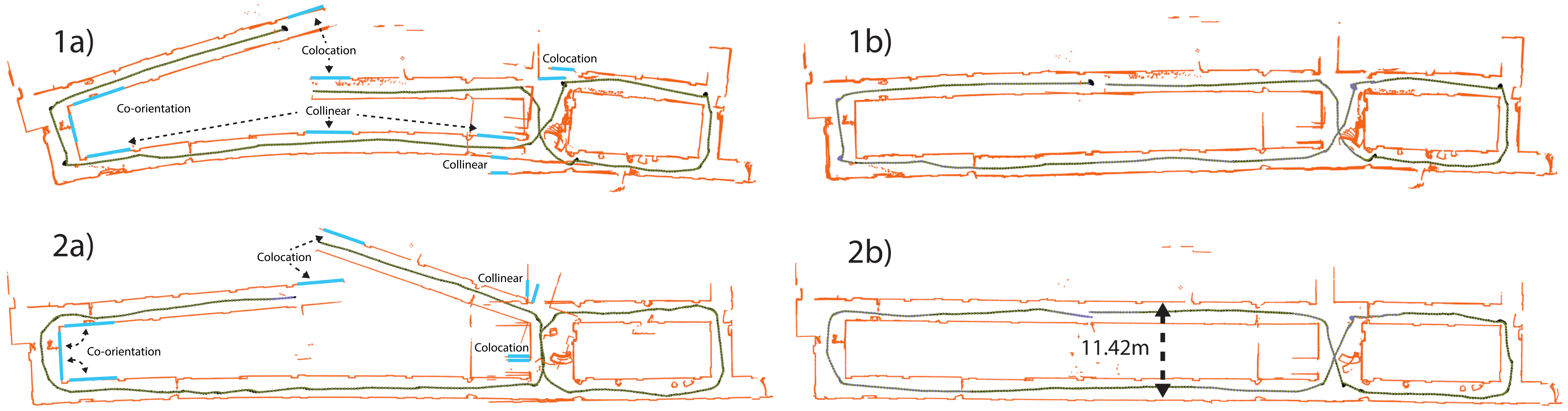}
	\caption{Initial and final maps from HitL-SLAM. Each map is of the same floor, and consists of between 600 and 700 poses. Maps in the left column (a) are initial maps, and maps in the right column (b) are final maps. Observations are shown in orange and poses are shown as arrows. Poses which are part of a human constraint are blue, while those which are not are in black.}
  \label{fig:F8}
\end{figure*}

\section{Results}

\begin{figure*}[t]
	\centering
	\includegraphics[scale = 0.048]{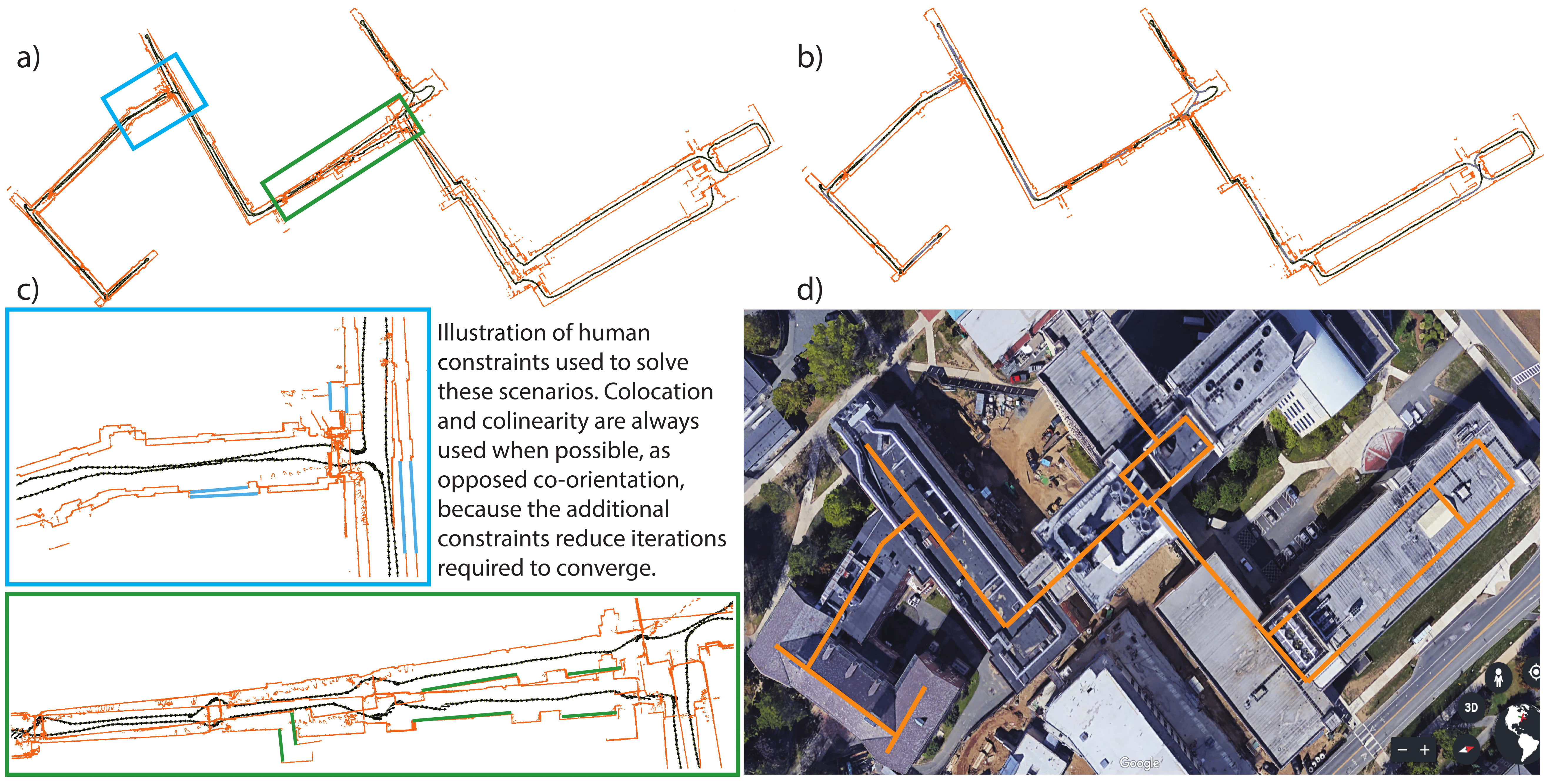}
	\caption{A large map a) corrected by HitL-SLAM b) using human correction, some of which are highlighted c). d) shows an approximate overlay of the map onto an aerial image of the complex from google earth. The map contains over 3000 poses.}
  \label{fig:BIG}
\end{figure*}

Evaluation of HitL-SLAM is carried out through two sets of experiments. The
first set is designed to test the accuracy of HitL-SLAM, and the second set is
designed to test the scalability of HitL-SLAM to large environments.

\begin{figure}[h!]
  \centering
  \includegraphics[scale=0.034]{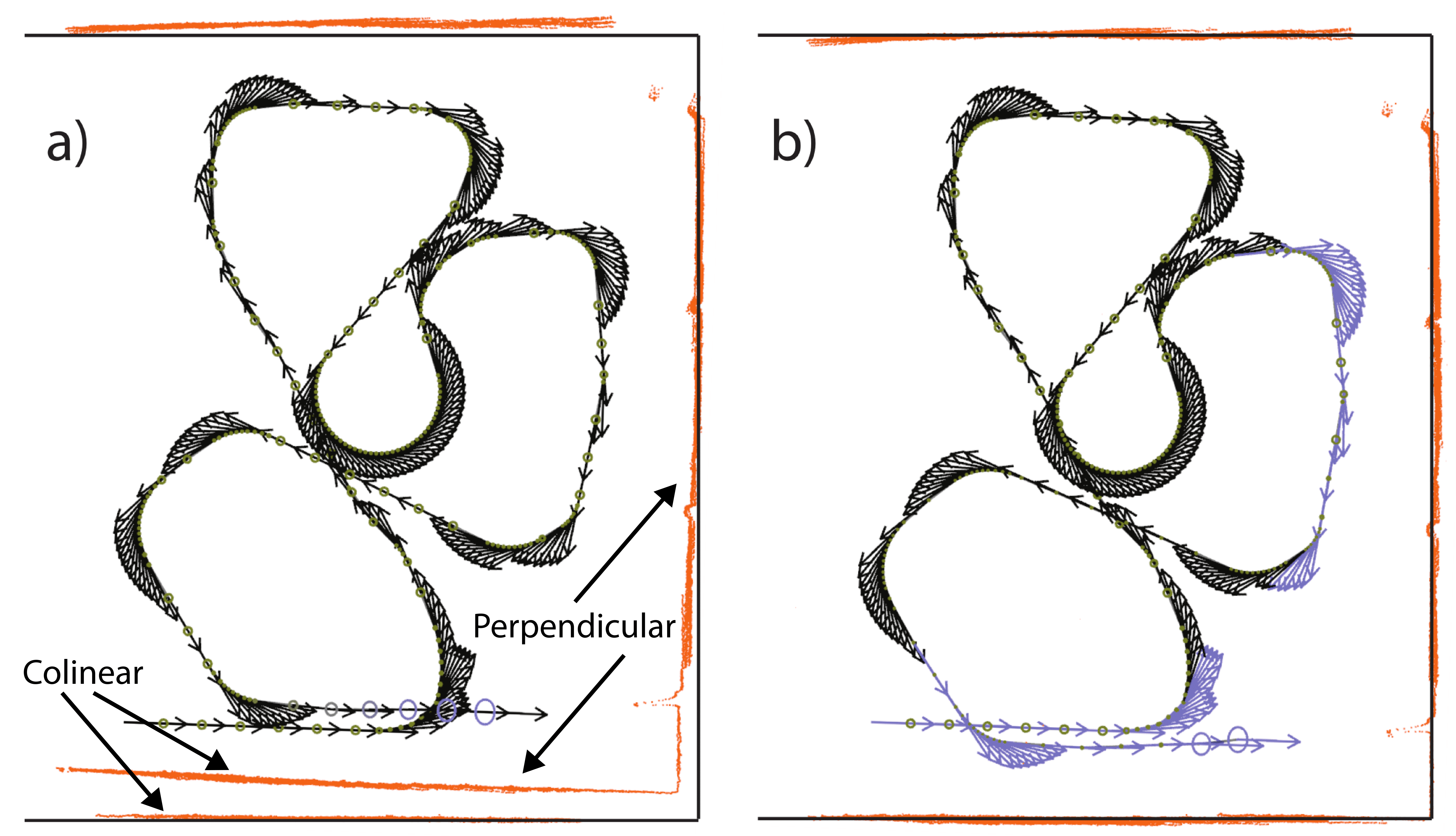}
  \caption{Initial a) and final b) maps for the `lost poses' experiment. Observations are shown in orange, poses are black arrows, and ground truth (walls) is represented by the black lines. Poses involved in human constraints are colored blue.}
\label{fig:LostPoses}
\end{figure}

To test the accuracy of HitL-SLAM, we construct a data set in a large
room during which no two parallel walls are simultaneously visible to
the robot. We do this by limiting the range of our robot's laser to 1.5m so that
it sees a wall only when very close. We then drive it around the room for which
we have ground truth dimensions. This creates sequences of ``lost" poses
throughout the pose-chain which rely purely on odometry to localize,
thus accruing error over time. We then impose human constraints on the resultant
map and compare to ground truth, shown in \figref{LostPoses}. Note that the
human corrections do not directly enforce any of the measured dimensions.
The initial map shows a room width of 5.97m, and an angle between opposite walls of $4.1^{\circ}$.
HitL-SLAM finds a room width of 6.31m, while the ground truth width is 6.33m, and 
produces opposite walls which are within $1^{\circ}$ of
parallel. Note also that due to the limited sensor range, the global
correctness must come from proper application of human constraints to the
factor graph including the ``lost" poses between wall observations.

To quantitatively evaluate accuracy on larger maps, where exact ground truth is sparse,
we measured an inter-corridor spacing (\figref{F8} 2b), which is constant
along the length of the building. We also measured angles between walls we know to be parallel or
perpendicular. The results for ground truth comparisons before and after HitL-SLAM, displayed in Table~\ref{table:gtcompare}, show that HitL-SLAM is able to drastically reduce map errors even when given poor quality initial maps.

\begin{figure}[h!]
  \centering
  \includegraphics[scale=0.086]{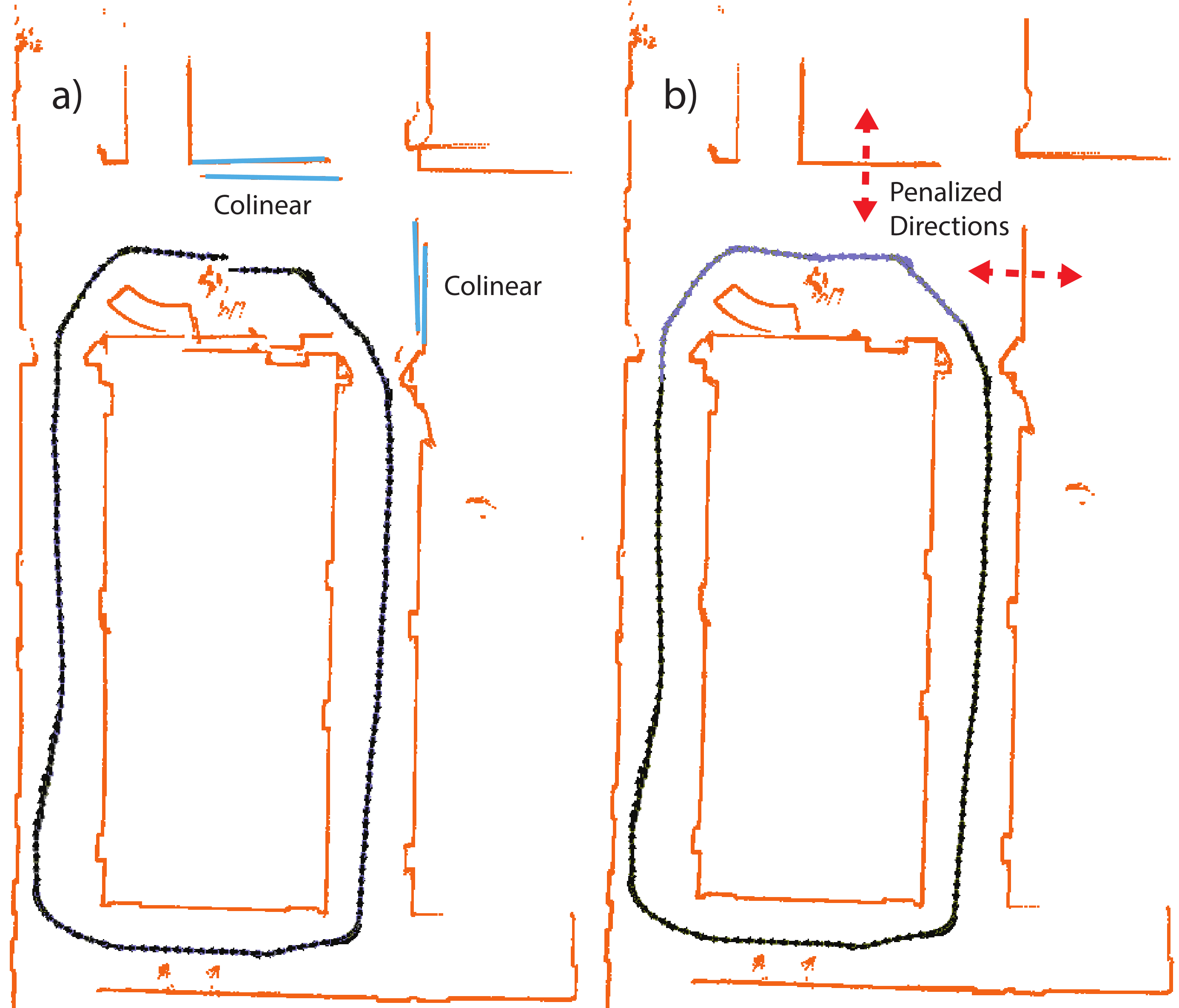}
  \caption{Example of how rank-deficient (collinear) constraints can be combined to effect a full rank constraint. Often, it is easier to tell if two sets of observations are collinear rather than colocated. Supporting rank-deficient constraints allows HitL-SLAM to operate on a larger set of maps.}
\label{fig:small_loop}
\end{figure}

We introduce an additional metric for quantitative map evaluation. We
define the pair-wise inconsistency $I_{i, j}$ between poses $x_i$ and $x_j$ to
be the area which observations from pose $x_i$ show as free space and
observations from pose $x_j$ show as occupied space. We define the total
inconsistency $\mathbf{I}$ over the map as the pair-wise sum of inconsistencies
between all pairs of poses, $\mathbf{I}=\sum_{i=1}^{N-1}\sum_{j=i+1}^N I_{i,j}$.
The inconsistency metric thus serves as a quantitative metric of global
registration error between observations in the map, and allows us to track the
effectiveness of global registration using HitL-SLAM over multiple iterations.
The initial inconsistency values for maps LGRC 3A and 3B were $297.5 m^2$ and
$184.3 m^2$, respectively. The final inconsistency values were $47.6 m^2$ and
$3.7 m^2$, respectively, for an average inconsistency reduction of $91 \%$
relative to the initial map, thus demonstrating the improved global consistency
of the map generated using HitL-SLAM.
\figref{F8} and \figref{small_loop} offer some qualitative examples of
HitL-SLAM's performance.

\begin{table}
  \centering
  \begin{tabular}{|l|c|c|c|c|c|c|}
    \hline
    \multirow{2}{*}{Map} & \multicolumn{2}{|c|}{Samples} & \multicolumn{2}{|c|}{Input
    Err.} &
    \multicolumn{2}{|c|}{{\small HitL-SLAM Err.}} \\
    \cline{2-7}
     & A & T & A($^\circ$) & T(m) & A($^\circ$) & T(m) \\
    \hline
    Lost Poses & 10 & 4 & 3.1 & 0.07 & 1.0 & 0.02 \\
    LGRC 3A & 14 & 10 & 9.8 & 3.3 & 1.5 & 0.06 \\
    LGRC 3B & 14 & 10 & 7.6 & 3.1 & 1.1 & 0.02 \\
    BIG MAP & 22 & 10 & 5.9 & 2.8 & 1.6 & 0.03 \\
    \hline
    Mean  & 60 & 34 & 6.74 & 2.71 & 1.4 & 0.04 \\
    \hline
  \end{tabular}
\caption{Quantitative mapping errors using HitL-SLAM compared to ground truth, in the
input maps, and after HitL-SLAM. The `Samples' column denotes how many pairwise feature comparisons were made on the map and then compared to hand-measured ground truth. Angular (A) errors are in degrees, translation (T) errors
in meters.}
\label{table:gtcompare}
\end{table}

To test the scalability of HitL-SLAM, we gathered several datasets with between 600 and 700 poses, and one with over 3000 poses and nearly 1km of indoor travel between three large buildings. \figref{F8} shows some of the moderately sized maps, and \figref{BIG} details the largest map. 16 constraints were required to fix the largest map, and computation time never exceeded the time required to re-display the map, or for the human to move to a new map location.

All maps shown in \figref{F8} were corrected interactively by the human using HitL-SLAM in under 15 minutes. Furthermore, HitL-SLAM solves two common problems which are difficult or impossible to solve via re-deployment: 1) a severely bent hallway, in \figref{F8} 1a), and 2) a sensor failure, in \figref{F8} 2a) which caused the robot to incorrectly estimate its heading by roughly 30 degrees at one point. Combined, these results show that incorporating human input into metric mapping can be done in a principled, computationally tractable manner, which allows us to fix metric mapping consistency errors in less time and with higher accuracy than previously possible, given a small amount of human input.

\section{Conclusion}

We present Human-in-the-Loop SLAM (HitL-SLAM), an algorithm designed to leverage
human ability and meta-knowledge as they relate to the data association problem
for robotic mapping. HitL-SLAM contributes a generalized framework for
interpreting human input using the EM algorithm, as well as a factor graph based
algorithm for incorporating human input into pose-graph SLAM. Future work in
this area could proceed towards further reducing the human requirements, and
extending this method for higher dimensional SLAM and for different sensor
types.

\section{Acknowledgments}

The authors would like to thank Manuela Veloso from
Carnegie Mellon University for providing the CoBot4 robot
used to collect the AMRL datasets, and to perform experiments
at University of Massachusetts Amherst. This work is supported in part by AFRL
and DARPA under agreement \#FA8750-16-2-0042.

\bibliography{HitL}
\bibliographystyle{aaai}

\end{document}